# Magnetic hysteresis in the microwave surface resistance of Nb samples in the critical state

M. Bonura, A. Agliolo Gallitto, and M. Li Vigni

CNISM and Dipartimento di Scienze Fisiche e Astronomiche, Università di Palermo, Via Archirafi 36, I-90123 Palermo (Italy)

**Abstract.** We discuss the hysteretic behavior of the field-induced variations of the microwave surface resistance in superconductors in the critical state. Measurements have been performed in a bulk sample of Nb and a powdered one at different values of the temperature. We discuss a model, based on the Coffey and Clem theory, in which we take into account the flux distribution inside the sample, due to the critical state. The experimental results are justified quantitatively in the framework of our model. We show that by fitting the experimental data it is possible to determine the value of the critical current density and its field dependence.

**PACS.** 74.25.Ha Magnetic properties – 74.25.Nf Response to electromagnetic fields (nuclear magnetic resonance, surface impedance, etc.) – 74.60.Ge Flux pinning, flux creep, and flux-line lattice dynamics

## 1 Introduction

Hysteretic behavior in the properties of several physical systems is a consequence of the fact that the system temporarily remains in a minimum of its free energy, if the thermodynamic equilibrium state has not been attained. So, hysteresis reveals the presence of metastable energy states. As it is well known, the magnetization of type-II superconductors exhibits hysteretic behavior because pinning centers hinder the fluxon motion [1]. The onset of the remanent magnetization, after applying and removing a magnetic field greater than the first penetration field, is a consequence of these effects. Magnetic hysteresis has been discussed for the first time by Bean who introduced the concept of the critical state of the fluxon lattice, characterized by a field-independent critical current density $J_c$ [2]. Successively, several models of critical state have been proposed, considering different field dependencies of $J_c$ [3–5]. The main consequence of the critical state is that, because of the effects of the pinning centers, the distribution of fluxons is not uniform inside the sample and the fluxon density is different for applied magnetic fields reached on increasing and decreasing values. The critical state is a metastable energy state and thermally activated processes allow fluxons surmounting the pinning barrier giving rise to an uniform flux distribution [1]. However, the relaxation times toward the thermal equilibrium state are in most cases much longer than the time during which the measurements are performed [1,6]. So, magnetic hysteresis can be detected in all the superconducting properties involving the presence of fluxons in the sample.

As it is well known [7–11], fluxon dynamics can be conveniently investigated by measuring the microwave (mw) surface resistance, $R_s$, which is proportional to the mw energy losses. Indeed, the variations of $R_s$, induced by magnetic fields higher than the first penetration field, are due to the presence and motion of fluxons within the mw-field penetration depth. The field-induced variations of $R_s$ in the mixed state have been studied by different authors [8,12,13]. Coffey and Clem (CC) have elaborated a comprehensive theory for the electromagnetic response of superconductors in the mixed state, in the framework of the two-fluid model of superconductivity [12]. The CC theory has been developed under the hypothesis that the local vortex magnetic field, $B(r)$, averaged over distances larger than several inter-vortex spacings, is uniform inside the sample; as a consequence, it neglects the effects of the critical state of the fluxon lattice. On the other hand, it is well known that the $H$-$T$ phase diagram of type-II superconductors is characterized by the presence of the irreversibility line, $H_{irr}(T)$, below which the magnetic properties of the superconducting sample become irreversible. The application of a DC magnetic field $H_0$ smaller than $H_{irr}(T)$ develops a critical state of the fluxon lattice; the magnetization of the sample shows a hysteretic behavior that brings about magnetic hysteresis in the $R_s(H_0)$ curves. Since the CC theory is inadequate to describe this effect, it has to be generalized by properly taking into account the flux distribution inside the sample.

Although different authors have discussed the hysteretic behavior of the $R_s(H_0)$ curves as due to the critical-state effects [14–16], to our knowledge a quantitative study has never been carried out. Very recently, we have proposed a method to take into account the flux distribution inside a superconducting sample in the critical state [17], in the framework of the CC theory. We have shown that the field dependence of the mw surface resistance is strongly affected by the specific profile of $B(r)$ determined by the field dependence of the critical current density, $J_c(B)$. So, in order to quantitatively justify the experimental results in superconductors in the critical state, the appropriate flux distribution in the sample has to be considerate. In this paper we discuss experimental results of field-induced variations of the mw surface resistance in bulk and powdered Nb samples. Measurements have been performed in the zero-field cooled samples, at different values of the temperature, by sweeping the DC magnetic field from zero up to a certain value, and back. At temperatures smaller enough than $T_c$, where pinning effects are significant, the $R_s(H_0)$ curve shows a hysteretic behavior that is related to the different flux distributions inside the sample, which arise on increasing and decreasing the external field. The experimental results have been justified quite well in the framework of our model by taking into account the proper flux distribution, due to the critical state.

## 2 Experimental Apparatus and Samples

The field-induced variations of $R_s$ have been studied in two samples of niobium, a bulk sample and a powdered one. The bulk sample has a nearly cylindrical shape, with radius $r_0 \approx 1.3$ mm and height $d \approx 3$ mm. The Nb powder has average grain size of $\sim 30$ $\mu$m and is placed in a Plexiglas holder of similar volume as the bulk sample.

The mw surface resistance is measured using the cavity-perturbation technique [18]. A copper cavity, of cylindrical shape with golden-plated walls, is tuned in the $TE_{011}$ mode resonating at $\omega/2\pi \approx 9.6$ GHz. The sample is located in the center of the cavity by a sapphire rod, in the region in which the mw magnetic field is maximum. The cavity is placed between the poles of an electromagnet which generates DC magnetic fields up to $H_0 \approx 1$ T. Two additional coils, independently fed, allow compensating the residual field and working at low magnetic fields. The sample and the field geometries are shown in Fig.1(a); the DC magnetic field is applied along the cylinder axis, the mw magnetic field, $H(\omega)$, is perpendicular to $H_0$. When the sample is in the mixed state, the induced mw current causes a tilt motion of the vortex lattice [19]; Fig.1(b) schematically shows the motion of a flux line.

The surface resistance of the sample is given by: $R_s = \Gamma(1/Q_L - 1/Q_U)$, where $Q_L$ is the quality factor of the cavity loaded with the sample, $Q_U$ that of the empty cavity and $\Gamma$ the geometric factor of the sample. The quality factor of the cavity is measured by means of an hp-8719D Network Analyzer.

## 3 Experimental Results

Fig.2 shows the temperature dependence of the surface resistance in the bulk and powdered Nb samples, at $H_0 = 0$. In order to disregard the geometrical factor, we have normalized the data to the value of the surface resistance in the normal state, $R_n$, at $T = T_c$. The bulk sample exhibits a narrow superconducting transition with onset $T_c \approx 9.2$ K. On the contrary, the powder shows a wide transition with onset $T_c \approx 8.0$ K; moreover, it exhibits a larger residual $R_s$, at the lowest temperature investigated, with respect to the bulk. These results suggest that the bulk sample is a high-quality Nb sample, while the powder sample is inhomogeneous.

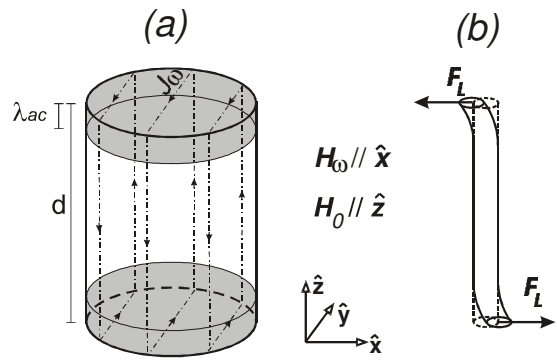

**Fig. 1.** (a) Current geometry in the sample surface. The shadowed areas indicate the sample regions where vortices experience the Lorentz force, $F_L$. (b) Schematic representation of the vortex motion.

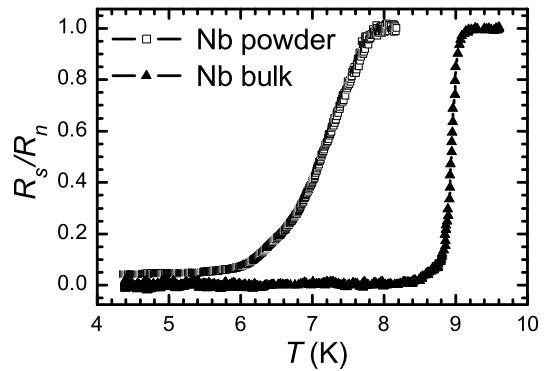

**Fig. 2.** Normalized values of the surface resistance, $R_s/R_n$, as a function of the temperature, obtained at $H_0 = 0$ for the two Nb samples. $R_n$ is the surface resistance at $T = T_c$.

The field induced variations of $R_s$ have been investigated for different values of the temperature. Each measurement has been performed by the following procedure: the sample has been cooled down to the desired value of the temperature in zero magnetic field (ZFC); the DC magnetic field has been increased up to a certain value and, successively, decreased down to zero. Fig.3 shows the field-induced variations of $R_s$ for the Nb bulk sample, obtained at different temperatures by sweeping $H_0$ from zero to the upper critical field, $H_{c2}(T)$, and back. In the figure, $\Delta R_s(H_0, T) \equiv R_s(H_0, T) - R_{res}$, where $R_{res}$ is the residual mw surface resistance at $T = 2.2$ K and $H_0 = 0$; moreover, the data are normalized to the maximum variation, $\Delta R_s^{max} \equiv R_n - R_{res}$. As one can see, $R_s$ does not show any variation as long as the magnetic field reaches a certain value, depending on $T$, that identifies the first-and, consequently, no hysteresis is detected. On decreasing the external field, the effects of the pinning centers become more and more important, and a critical state of the fluxon lattice develops. At low temperatures, the hysteresis is evident in a wide range of magnetic fields; on increasing the temperature, the pinning weakens and the field range in which the hysteresis is present shrinks.

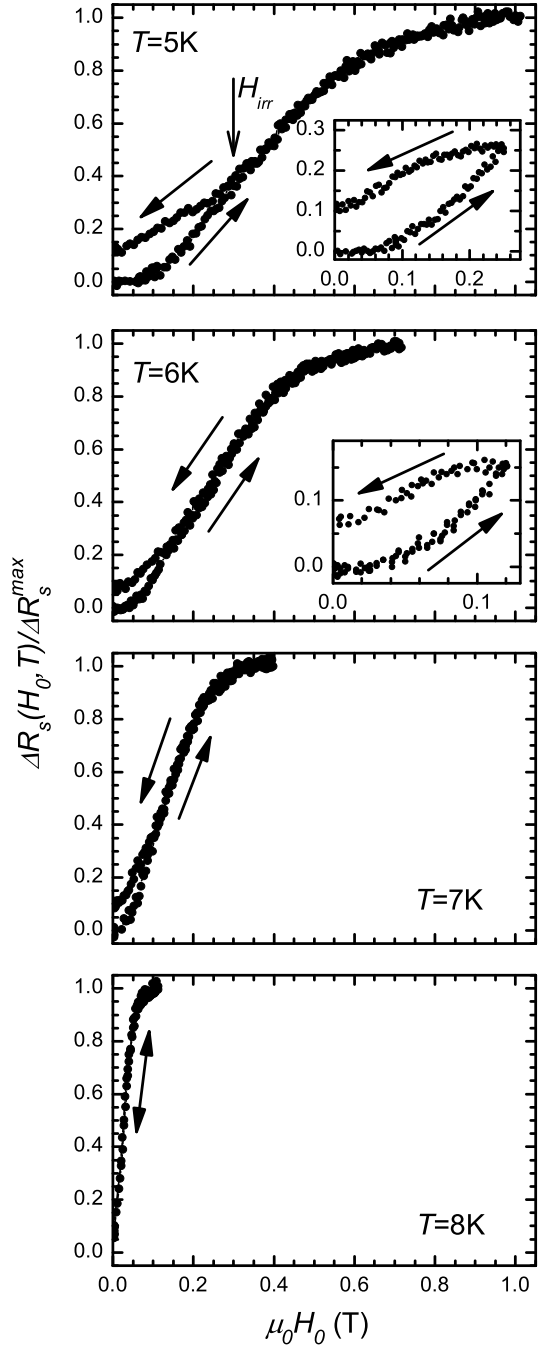

**Fig. 3.** Field-induced variations of $R_s$ for the Nb bulk sample, at different temperatures. $\Delta R_s(H_0, T) \equiv R_s(H_0, T) - R_{res}$, where $R_{res}$ is the residual mw surface resistance at $T = 2.2$ K and $H_0 = 0$; $\Delta R_s^{max} \equiv R_n - R_{res}$.

From isothermal $R_s$ vs. $H_0$ curves, obtained at different temperatures, we have deduced the temperature dependence of $H_p$, $H_{irr}$ and $H_{c2}$; the values for the bulk sample are shown in Fig.4. The line in plot (a) has been obtained by fitting the experimental data with the law $H_p(T) = H_p(0)[1 - (T/T_c)^\beta]$; we have obtained, as best-fit parameters, $H_p(0) = 750 \pm 40$ mT and $\beta = 2.3 \pm 0.3$. The deduced values of the characteristic fields are consistent with results reported in the literature for Nb superconductor [20–23]. In particular, the value of $H_p(0)$ is consistent with the reported values of the lower magnetic field; so, we assume that surface barrier [23] and/or demagnetization effects can be neglected in this sample.

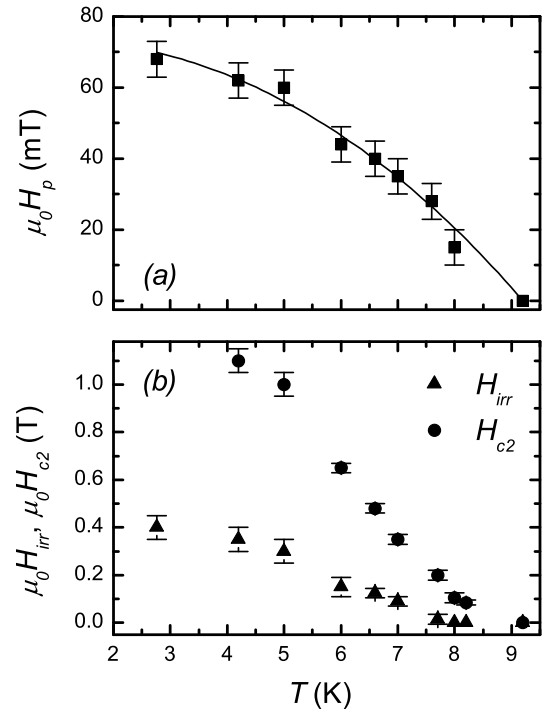

**Fig. 4.** Temperature dependence of the characteristic fields, $H_p$, $H_{irr}$ and $H_{c2}$, of the bulk sample, deduced by $R_s(H_0)$ measurements.

penetration field, $H_p$. For $H_0 > H_p$, vortices start to penetrate the sample and, consequently, $R_s$ increases.

Up to about 2 K below $T_c$, the curves of Fig.3 exhibit a magnetic hysteresis that disappears for $H_0$ higher than a certain value, which is indicated in the figure by $H_{irr}$. The results can be qualitatively understood considering that at fields near $H_{c2}(T)$ the pinning centers are ineffective, the fluxons distribute uniformly in the sample

Fig.5 shows the normalized field variations of the surface resistance for the Nb-powder sample, at different values of the temperature. Since in this sample the hysteresis amplitude is smaller than in the bulk, the results are shown in an enlarged scale (and consequently in a restricted field range) to see it clearly. Similar to what occurs in the Nb bulk, the hysteresis shrinks on increasing the temperature; eventually, it disappears for $T > 6$ K. A detailed analysis of the results obtained at low magnetic fields shows that in this sample the surface resistance exhibits an initial slow variation followed by a faster one. The initial increase starts at $H_0$ values too small ($< 10$ mT at $T = 2.2$ K) to be ascribed to penetration of Abrikosov vortices. We suggest that the slow variation is due to the presence of weak links that give an inter-grain contribution to energy losses at fields lower than $H_{c1}$. This hypothesis is corroborated by the fact that in this sample we have detected mw-second-harmonic emission, clearly due to the presence of weak links [24]; it is also consistent with the results of Fig.2, which show a large residual resistance and a broad superconducting transition.

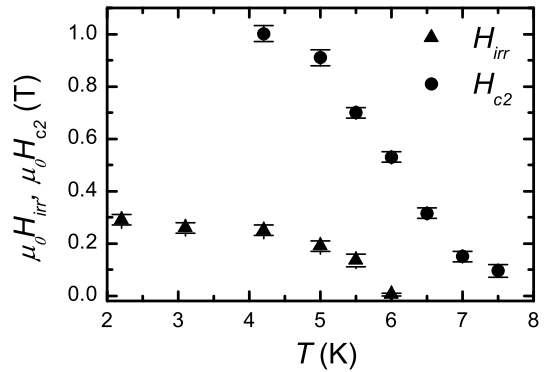

**Fig. 6.** Temperature dependence of the characteristic fields, $H_{irr}$ and $H_{c2}$, of the Nb powder, deduced by $R_s(H_0)$ measurements.

## 4 The model

In the London local limit, the surface resistance is proportional to the imaginary part of the complex penetration depth, $\widetilde{\lambda}$, of the em field:

$$R_s = -\mu_0 \omega \ \text{Im}[\widetilde{\lambda}(\omega, B, T)]. \qquad (1)$$

Furthermore, we would recall that the real part of $\widetilde{\lambda}$ defines the ac penetration depth, $\lambda_{ac}$.

Coffey and Clem have elaborated a comprehensive theory for the electromagnetic response of superconductors in the mixed state, by taking into account flux flow, flux creep and flux pinning, in the framework of the two-fluid model of superconductivity [12]. The theory has been developed under two basic assumptions: i) inter-vortex spacing much less than the field penetration depth; ii) uniform vortex distribution in the sample. With these assumptions, vortices generate a magnetic induction field, $B$, uniform in the sample.

In the CC model, $\widetilde{\lambda}(\omega, B, T)$ is given by

$$\widetilde{\lambda}(\omega, B, T) = \sqrt{\frac{\lambda^2(B,T) + (i/2)\widetilde{\delta}_v^2(\omega,B,T)}{1 - 2i\lambda^2(B,T)/\delta_{nf}^2(\omega,B,T)}}, \qquad (2)$$

with

$$\lambda(B,T) = \frac{\lambda_0}{\sqrt{[1-(T/T_c)^4][1-B/B_{c2}(T)]}}, \qquad (3)$$

$$\delta_{nf}(\omega, B, T) = \frac{\delta_0}{\sqrt{1-[1-(T/T_c)^4][1-B/B_{c2}(T)]}}, \qquad (4)$$

where $\lambda_0$ is the London penetration depth at $T = 0$ and $\delta_0$ is the normal-fluid skin depth at $T = T_c$.

$\widetilde{\delta}_v$ is the effective complex skin depth arising from the vortex motion; it depends on the relative magnitude of the viscous and restoring-pinning forces. $\widetilde{\delta}_v$ can be written in terms of two characteristic lengths, $\delta_f$ and $\lambda_c$, arising from the contributions of the viscous and the restoring-pinning forces, respectively:

$$\frac{1}{\widetilde{\delta}_v^2} = \frac{1}{\lambda_c^2} - \frac{2i}{\delta_f^2} \qquad (5)$$

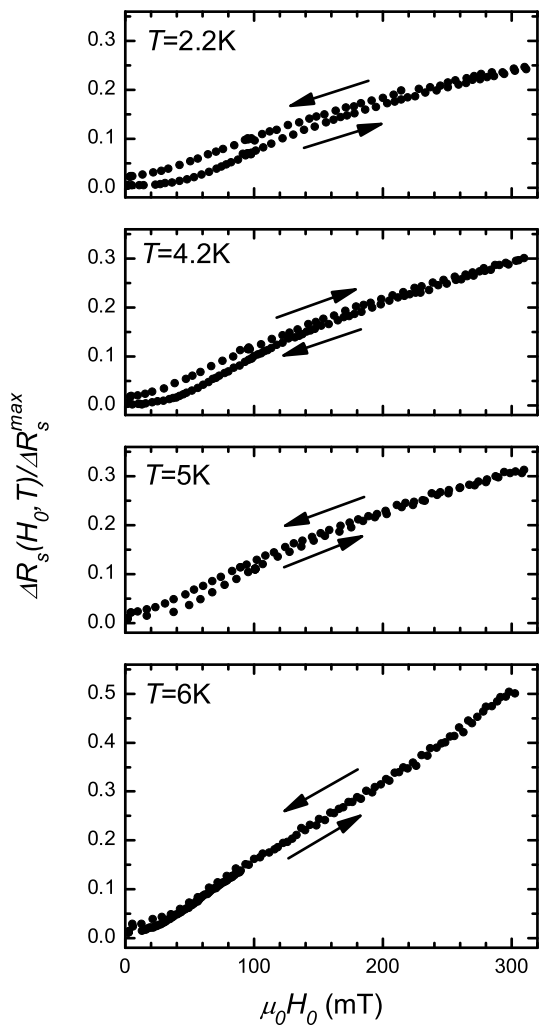

**Fig. 5.** Field-induced variations of $R_s$, for the Nb-powder sample, at different temperatures. The data are normalized to the maximum variation, $\Delta R_s^{max} \equiv R_n - R_{res}$.

In Fig.6 we report the temperature dependence of the characteristic fields, $H_{irr}$ and $H_{c2}$, deduced from the isothermal $R_s$ vs. $H_0$ curves, for the powder of Nb. Both $H_{irr}$ and $H_{c2}$ of this sample are smaller than those of the bulk sample.

where

$$\lambda_c^2 = \frac{B\phi_0}{\mu_0 k_p}, \quad (6)$$

$$\delta_f^2 = \frac{2B\phi_0}{\mu_0 \omega \eta}, \quad (7)$$

with $k_p$ the restoring-force coefficient, $\eta$ the viscous-drag coefficient and $\phi_0$ the quantum of flux.

The effectiveness of the two terms in equation (5) depends on the ratio $\omega_0 = k_p/\eta$, which defines the depinning frequency [7]. When the frequency of the em wave, $\omega$, is much lower than $\omega_0$, the motion of fluxons is ruled by the restoring-pinning force. On the contrary, for $\omega \gg \omega_0$, the contribution of the viscous-drag force predominates and the induced em current makes fluxons move in the flux-flow regime. In this case, in fact, the flux-line motion takes place around the minimum of the pinning potential well and, consequently, the restoring-pinning force is nearly ineffective.

As it is clear from Eqs.(1-4), it is expected that the features of the $R_s(H_0)$ curves strongly depend on the applied-field dependence of $B$. On the other hand, the CC theory is strictly valid when $B$ is uniform inside the sample; in particular for $H_0 \gg H_{c1}$, the $R_s(H_0)$ curves can be described from Eqs.(1-4) setting $B = \mu_0 H_0$. When the sample is exposed to a DC field smaller than $H_{irr}(T)$, the assumption of uniform $B$ is not longer valid and the CC theory does not correctly describe the field-induced variations of $R_s$ [17].

Since the energy losses occur within the mw-field penetration depth $\lambda_{ac}$, an important parameter to determine in what extent the non-uniform $B$ distribution affects the $R_s(H_0)$ curves is the variation of the local magnetic induction within $\lambda_{ac}$. So, it is expected that the disagreement between the results expected by the CC theory and the experimental results is particularly evident when the DC field is perpendicular to the mw magnetic field. Indeed, in this case, all the fluxons present in the sample experience the Lorentz force due to the mw current and the whole vortex lattice undergoes a tilt motion [19]. Recently, we have investigate just this case [17]; by considering the distribution of $B$ due to the critical state, in the framework of the CC model, we have shown that the the parameters that more affect the $R_s(H_0)$ curves are the full penetration field, $H^*$, and the field dependence of $J_c$.

When $B$ is not spatially uniform, the behavior of $R_s(H_0)$ differs from that expected for uniform B because different regions of the sample contribute to the energy losses in a different extent, dependently on the local $B$ value in each region. However, in order to take into account the $B$ distribution, one can imagine the sample surface as subdivided in different regions in such a way that $B$ is locally uniform in each of them. The energy losses of the whole sample are determined by the surface-resistance contribution of each region, that depends on the local $B$ value by Eqs. (1-4). The measured surface resistance is an averaged value over the whole sample:

$$R_s = \frac{1}{S} \int_\Sigma R_s(|B(\boldsymbol{r})|)\, dS, \quad (8)$$

where $\Sigma$ is the sample surface, $S$ is its area and $\boldsymbol{r}$ identifies the surface element.

We would remark that, in type-II superconductors in the mixed state, the energy losses are due to both the presence of fluxons, which bring about normal fluid in their cores, and their motion. However, the pinning effects are particularly enhanced at temperature smaller enough than $T_c$, where the dissipations are essentially due to vortex motion. So, the main contribution to $R_s$ comes from the sample regions in which fluxons experience the Lorentz force due to the mw current, i.e. where $\boldsymbol{H}_0 \times \boldsymbol{J}_\omega \neq 0$.

As already mentioned, one of the main consequences of the critical state consists in a different $B$ distribution when the applied magnetic field is reached at increasing or decreasing values. This is responsible for the appearance of magnetic hysteresis in the $R_s(H_0)$ curves. In the following of this section we report some expected results on the hysteretic behavior of $R_s(H_0)$.

In order to calculate the normalized values of the surface resistance, using Eqs.(1-4), it is necessary to know the ratio $\lambda_0/\delta_0$, the lower and upper critical fields and the depinning frequency. Moreover, to take into account the critical-state effects by Eq.(8), it is also essential to know the $B$ profile inside the sample, determined by $J_c(B)$. For simplicity, we consider a sample of cylindrical shape, with the DC magnetic field applied along its axis, which is in a critical state à la Bean, i.e. $J_c = J_{c0}$ independent of $B$ (we recall that in this case the full-penetration field is given by $H^* = J_{c0}r_0$, being $r_0$ the cylinder radius). Furthermore, we suppose that fluxons move in the flux-flow regime, i.e. $\omega \gg \omega_0$. In this case, considering the expression of the viscous coefficient proposed by Bardeen and Stephen [25], it results $\widetilde{\delta}_v^2 = \delta_0^2 B/B_{c2}(T)$. We would remark that the analysis can be easily extended to a more general case provided that the field dependencies of $J_c$ and $\omega_0$ are known.

Fig.7 shows the expected $R_s(H_0)/R_n$ curves for two different values of the full-penetration field, $H^*$, obtained varying the DC magnetic field in the range $0 \to H_{max} \to 0$, with $H_{max} = H_{c2}/6$. The curves have been obtained supposing that the sample is in a critical state à la Bean and the fluxons move in the flux-flow regime. The calculations have been performed with $\lambda_0/\delta_0 = 0.025$, $T = T_c/3$, $H_{c1} = 0$.

The curves of Fig.7 describe the expected results in two samples having the same properties but different size, or alternatively the same size but different critical current. As one can see, both the curves show a counterclockwise hysteresis, but they exhibit different peculiarities. The most evident difference concerns the amplitude of the hysteresis loop, which is larger when $H^*$ and $H_{max}$ are of comparable magnitude. Another property that characterizes the $R_s(H_0)$ curves is the concavity of the increasing-field branch, which is different for curves (a) and (b); in particular, the increasing-field branch of curve (b) shows a change of concavity when $H_0$ overcomes $H^*$. On the contrary, the concavity of the decreasing-field branch is negative in both the cases. All these peculiarities can be qualitatively understood by looking at Fig.8, which shows the $B$ profile in two samples, in a critical state à la Bean, having the same critical current but different size and, consequently, different $H^*$.

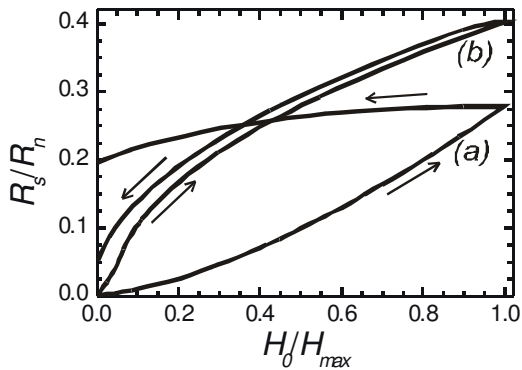

**Fig. 7.** Expected field-induced variations of $R_s$ for two different values of $H^*$: (a) $H^* = H_{max}$; (b) $H^* = 0.05 H_{max}$. The expected results have been obtained as described in the text using $H_{max} = H_{c2}/6$, $H_{c1} = 0$, $T = T_c/3$, $\lambda_0/\delta_0 = 0.025$.

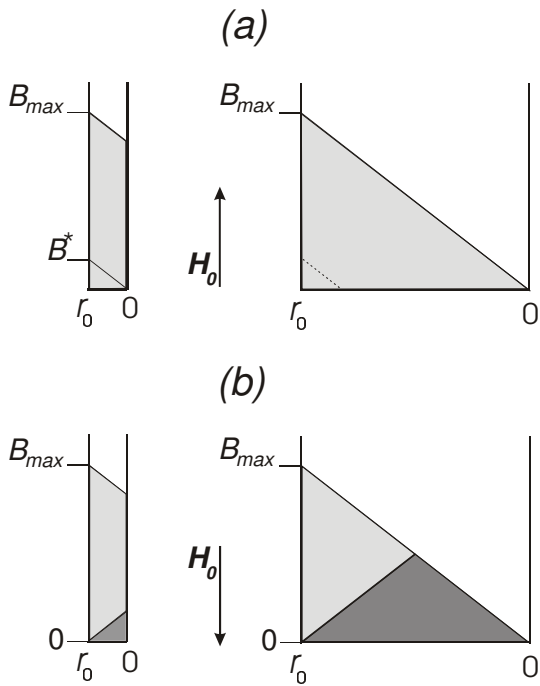

**Fig. 8.** Magnetic-induction profiles in ZFC superconducting cylinders in the critical state à la Bean, for different values of $H^*$ due to their different sizes. (a): $B$ profiles obtained after $H_0$ was increased from 0 to $H_{max}$. (b): $B$ profiles when $H_0$ was swept in the range $0 \rightarrow H_{max} \rightarrow 0$; dark-shadowed areas represent the remanet magnetic induction.

In Fig.7, the change of concavity in the increasing-field branch of curve (b) is ascribable to the change of the external field dependence of $B$ occurring when $H_0$ reaches $H^*$. In particular, by looking at Fig.8 (a) one can see that for $H_0 < H^*$ on increasing the external field from zero up to $H^*$ more and more regions contribute to the mw losses, this gives rise to a positive concavity of the $R_s(H_0)$ curve. For $H_0 > H^*$, in the whole sample the local magnetic induction linearly depends on the external field giving rise to a negative concavity of the $R_s(H_0)$ curve. The negative concavity of the decreasing-field branches of the $R_s(H_0)$ curves is a consequence of the shape of the magnetization hysteresis; indeed, soon after the field-sweep direction is reversed, the trapped flux does not appreciably change, giving rise to an initial plateau in the decreasing-field branches of the $R_s(H_0)$ curves.

The different amplitudes of the hysteresis loop of curves (a) and (b) of Fig.7 can be qualitatively understood looking at Fig.8, which shows the average magnetic induction over the sample at $H_0 = H_{max}$ (gray-shadowed areas in panel $(a)$) and the remanent magnetic induction after the complete cycle of $H_0$ (dark-shadowed areas in panel $(b)$). As one can see, the lower $H^*$ the lower the ratio between the dark and gray areas; therefore, the largest hysteresis comes out when $H^* \sim H_{max}$.

## 5 Discussion

Very recently, we have investigated the field-induced variations of $R_s$ at increasing magnetic fields in a bulk-Nb sample cut from the same batch from which the bulk sample here investigated has been extracted [17]. We have shown that in the whole field range $H_{c1} \leq H_0 \leq H_{c2}$ the experimental data are quantitatively justified in the framework of the model reported in Sec.3. Since the value of the depinning frequency reported in the literature for high-quality Nb samples are much lower than the working frequency [26], we assumed that fluxons move in the flux-flow regime. The best fit of the data for increasing fields was obtained by setting $\lambda_0/\delta_0 = 3 \times 10^{-2}$ [27] and using a linear field dependence of $J_c$ at low fields followed by an exponential decrease [17]. All these values of the parameters are consistent with those reported in the literature for Nb [21,27,28].

In this paper, we discuss the hysteretic behavior of the $R_s(H_0)$ curves. A comparison between Fig.3 and Fig.5 shows that the amplitude of the hysteresis is larger in the Nb bulk than in the Nb powder. This finding qualitatively agrees with the results reported in Fig.7, which shows the expected $R_s(H_0)$ curves for samples in the critical state à la Bean. However, the experimental results here reported cannot be fully justified using a field-independent $J_c$ in a wide range of magnetic fields.

The hysteretic behavior observed in the bulk sample can be fully justified in the framework of the model discussed in Sec.3. We have fitted the experimental results of the minor loops shown in the insets of Fig.3, in which the hysteresis is enough large to be detected with a good resolution. The best-fit curves are shown in Fig.9, they have been obtained using for $H_p(T)$ and $H_{c2}(T)$ the values of Fig.4, letting them vary within the experimental accuracy. According to the results reported in Ref.[17], we have used $\lambda_0/\delta_0 = 3 \times 10^{-2}$, the linear dependence of the critical current density $J_c(B) = J_{c0} - \alpha B$, and we have set the induction field at the edge of the sample $B(r_0) = \mu_0(H_0 - H_p)$.

The values of the parameters which best fit the results obtained at $T = 5$ K are: $\mu_0 H_p = 62$ mT, $\mu_0 H_{c2} = 1.05$ T, $J_{c0} = 12 \times 10^7$ A m$^{-2}$ and $\alpha = 1.75 \times 10^7$ A m$^{-2}$T$^{-1}$; while those which best fit the results obtained at $T = 6$ K are: $\mu_0 H_p = 45$ mT, $\mu_0 H_{c2} = 0.65$ T, $J_{c0} = 7.5 \times 10^7$ A m$^{-2}$ and $\alpha = 7.5 \times 10^7$ A m$^{-2}$T$^{-1}$. Although the values of $J_c$ reported in the literature for Nb spread over a wide range, the $J_{c0}$ values we found are consistent with those obtained in high-quality Nb samples [21]. The decrease of $J_{c0}$, as well as the growth of $\alpha$, on increasing the temperature are ascribable to the weakening of the pinning.

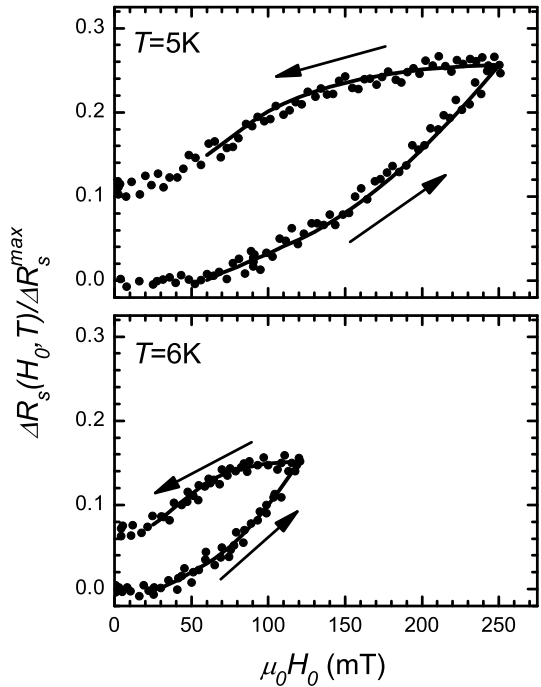

**Fig. 9.** Best-fit curves of the experimental data for the Nb bulk sample, obtained as explained in the text. The expected curve for $T = 5$ K has been obtained using: $\lambda_0/\delta_0 = 3 \times 10^{-2}$, $\mu_0 H_p = 62$ mT, $\mu_0 H_{c2} = 1.05$ T, $J_{c0} = 12 \times 10^7$ A m$^{-2}$ and $\alpha = 1.75 \times 10^7$ A m$^{-2}$T$^{-1}$. The expected curve for $T = 6$ K has been obtained using: $\lambda_0/\delta_0 = 3 \times 10^{-2}$, $\mu_0 H_p = 45$ mT, $\mu_0 H_{c2} = 0.65$ T, $J_{c0} = 7.5 \times 10^7$ A m$^{-2}$ and $\alpha = 7.5 \times 10^7$ A m$^{-2}$T$^{-1}$.

A quantitative study of the field-induced variations of $R_s$ in the Nb powder is difficult to carry out for several reasons. Firstly, as it has been already mentioned, the weak-link effect on the energy losses hinders to measure the magnetic field at which Abrikosov fluxons penetrate the powder grains; so, $H_p$ has to be taken as fitting parameter. Moreover, we have tried to fit the experimental data supposing fluxons move in the flux-flow regime, but we did not get good results. After several fitting attempts, it came out that in this sample the depinning frequency is of the same order of the working frequency and that it cannot be assumed independent of the magnetic field. For all these reasons, we have fitted the experimental results obtained at $T = 4.2$ K in the restricted range of magnetic fields $70 \div 150$ mT. At this temperature, the amplitude of the hysteresis loop is large enough to be detected with a good resolution and further the upper critical field can be experimentally deduced (see Fig. 6). Moreover, in order to reduce the number of free parameters, we have assumed that, in this restricted range of magnetic fields, the critical current density does not depend on $B$, setting $J_c = J_{c0}$, and we have used the same value of $\lambda_0/\delta_0$ of the bulk sample. In this way, the free fitting parameters are $H_p$, $J_{c0}$ and $\omega_0(B)$, which has been set as $\omega_0(B) = 2\pi(\nu_0 - \gamma B)$. The best-fit curve is shown in Fig.10, it has been obtained with $\mu_0 H_p = 25$ mT, $\mu_0 H_{c2} = 1.0$ T, $J_{c0} = 5 \times 10^8$ A m$^{-2}$, $\nu_0 = 5 \times 10^9$ s$^{-1}$ and $\gamma = 11.4 \times 10^9$ s$^{-1}$T$^{-1}$.

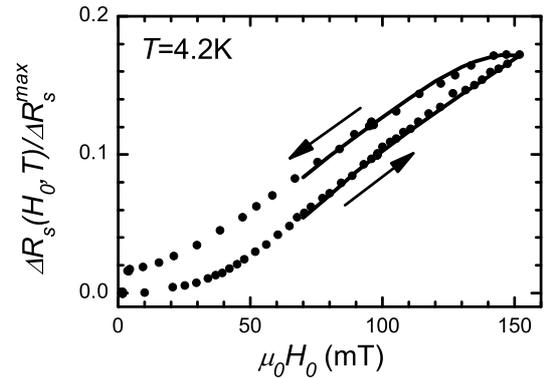

**Fig. 10.** Best-fit curve of the experimental data for the Nb powder, obtained as explained in the text. The expected curve has been obtained using: $\lambda_0/\delta_0 = 3 \times 10^{-2}$, $\mu_0 H_p = 25$ mT, $\mu_0 H_{c2} = 1.0$ T, $J_{c0} = 5 \times 10^8$ A m$^{-2}$, $\nu_0 = 5 \times 10^9$ s$^{-1}$ and $\gamma = 11.4 \times 10^9$ s$^{-1}$T$^{-1}$

As one can note, the value of the critical current density we found in the powder is higher than that of the bulk sample; this finding is consistent with the higher value of the depinning frequency for the powder with respect to the bulk. Indeed, our results show that the depinning frequency in the powder sample is roughly $\omega/2$, while in the bulk it results $\omega_0/\omega \ll 1$. It is worth to remark that the value of $J_c$ we found refers to an average value of the critical current density of the powder grains. So, our results suggest that the pinning is weaker in the bulk sample than in the grains of the powder sample. Nevertheless, the amplitude of the hysteresis loop is smaller in the powder than in the bulk; this is due to the reduced value of the full-penetration field of the powder grains.

## 6 Conclusions

We have discussed, both experimentally and theoretically, the field-induced variations of the microwave surface resistance in two Nb samples in the critical state. The measurements have been performed in a bulk sample and in a powdered one, at different temperatures, by sweeping the DC magnetic field from zero to a certain value, and back. At temperatures smaller enough than $T_c$, where the pinning is significant, the $R_s(H_0)$ curves show a magnetic hysteresis that is related to the different flux distributions for DC magnetic field reached at increasing and decreasing values. The experimental results have been quantitatively justified by a model, based on the Coffey and Clem theory, in which we take into account the distribution of the induction field inside the sample. To our knowledge, for the first time we have quantitatively discussed the hysteretic behavior of the mw surface resistance.

In the bulk sample, the experimental results have been quite well justified in the whole range of magnetic fields in which the hysteresis has been detected. Considering the value of the depinning frequency reported in the literature for high-quality Nb samples, we have fitted the experimental data supposing that the mw current induces fluxons to move in the flux-flow regime and taking the critical current density and its field variation as fitting parameters. The values of $J_c$ we found, as well as its field dependence, are consistent with those reported in the literature, showing that measurements of field-induced variations of the surface resistance provide a method to determine the critical current density, which is alternative to the magnetization measurements.

In the powder sample, in order to use a reduced number of free parameters, we have fitted the experimental data in a restricted range of magnetic fields. From the fitting results, higher values of both the critical current density and the depinning frequency come out, suggesting that the pinning is stronger in the powder grains than in the bulk sample. Nevertheless, in the powder we have observed a smaller hysteresis because the main parameter affecting the hysteresis amplitude is the full penetration field, as it similarly occurs in the magnetization curve.

# 7 Acknowledgements

The authors are very glad to thank E. Di Gennaro for his interest and helpful suggestions; G. Lapis and G. Napoli for technical assistance.

July 28, 2006